\documentclass[a4paper,fleqn,usenatbib]{mnras}

\usepackage[T1]{fontenc}
\usepackage{ae,aecompl}

\usepackage{graphicx}	% Including figure files
\usepackage{amsmath}	% Advanced maths commands
\usepackage{amssymb}	% Extra maths symbols

\hypersetup{breaklinks}
\hypersetup{pdfauthor={Mathias Zechmeister},
            pdftitle={Solving Kepler's equation with CORDIC double iterations}}
\usepackage{breakurl}  % needed for ps2pdf, not for dvipdfm

\usepackage[all]{hypcap} % center targets after click

\newcommand{\ZeXVIII}[1][]{\href{https://ui.adsabs.harvard.edu/abs/2018A\%26A...619A.128Z}{Ze18#1}}
\renewcommand{\dagger}{^{\star\star}}

\title[Solving KE with CORDIC double iterations]{Solving Kepler's equation with CORDIC double iterations\thanks{\protect\label{fn:thanks}Code available at \protect\url{https://github.com/mzechmeister/ke/}.}}

\author[M.~Zechmeister]{
M.~Zechmeister\thanks{E-mail: zechmeister@astro.physik.uni-goettingen.de}
\\
\\
Institut f\"ur Astrophysik, Georg-August-Universit\"at, Friedrich-Hund-Platz
1, 37077 G\"ottingen, Germany
}

\date{Accepted 2020 August 6. Received 2020 August 5; in original form 2020 May 11.}

\pubyear{2020}

\makeatletter

\providecommand{\tabularnewline}{\\}

\usepackage{txfonts}
\bibpunct{(}{)}{;}{a}{}{,}  % to follow the A&A style
\@ifundefined{lyxadded}{}{
   \hypersetup{citecolor = green}
}
\usepackage{color}

\usepackage{listings}
\usepackage[scaled=0.82]{beramono}  % for the code lstlisting

\definecolor{deepblue}{rgb}{0,0,0.5}
\definecolor{deepred}{rgb}{0.6,0,0}
\definecolor{deepgreen}{rgb}{0,0.5,0} 

\DeclareMathOperator{\sign}{sign}

\makeatletter

\lst@AddToHook{EveryPar}{%
    \edef\@temp{\noexpand\label{\lst@label-\arabic{lstnumber}}}%
    \@temp
}

\newcommand\footnoteref[1]{\protected@xdef\@thefnmark{\ref{#1}}\@footnotemark} % to reference the thanks footnote

\newcommand*\tablefootmark[1]{%
  \unskip
  \hbox{\@textsuperscript{\normalfont\itshape\ignorespaces#1}}%
  \,%
  \ignorespaces
}
\newcommand\tablefoottext[2]{%
  \hbox{\@textsuperscript{\normalfont({\itshape\ignorespaces#1})}}%
  ~%
  \ignorespaces
  #2\ \ignorespaces%
}

\makeatother

\let\orgautoref\autoref
\newcommand{\hautoref}{%
         \def\figureautorefname{Fig.}%
         \def\sectionautorefname{Sect.}%
         \def\subsectionautorefname{Sect.}%
         \def\subsubsectionautorefname{Sect.}%
         \orgautoref
}

\renewcommand{\autoref}[1]{%
    \def\equationautorefname~##1\null{Eq.~(##1)\null}%
    \def\lstnumberautorefname~##1\null{L\,##1\null}% small separator
    \hautoref{#1}%
}

\newcommand{\hAutoref}{% Capital at sentence start.
         \def\figureautorefname{Figure}%
         \def\sectionautorefname{Section}%
         \def\subsectionautorefname{Section}%
         \def\subsubsectionautorefname{Section}%
         \orgautoref}

\newcommand{\Autoref}[1]{%
    \def\equationautorefname~##1\null{Equation~(##1)\null}%
    \hAutoref{#1}%
}

\everymath{
\medmuskip=0.5\medmuskip
\thickmuskip=0.5\thickmuskip
}

\usepackage{chngcntr}
\AtBeginDocument{\counterwithin{lstlisting}{section}}

\DeclareMathAlphabet{\pazocal}{OMS}{zplm}{m}{n}
\newcommand{\unif}{\pazocal{U}}

\makeatother

\usepackage{listings}
\renewcommand{\lstlistingname}{Listing}

\begin{document}
\label{firstpage}
\pagerange{\pageref{firstpage}--\pageref{lastpage}}
\maketitle

\begin{abstract}
In a previous work, we developed the idea to solve Kepler\textquoteright s
equation with a CORDIC-like algorithm, which does not require any
division, but still multiplications in each iteration. Here we
overcome this major shortcoming and solve Kepler\textquoteright s
equation using only bitshifts, additions, and one initial multiplication.
We prescale the initial vector with the eccentricity and the scale
correction factor. The rotation direction is decided without correction
for the changing scale. We find that double CORDIC iterations are
self-correcting and compensate possible wrong rotations in subsequent
iterations. The algorithm needs 75\% more iterations and delivers
the eccentric anomaly and its sine and cosine terms times the eccentricity.
The algorithm can be adopted for the hyperbolic case, too. The
new shift-and-add algorithm brings Kepler's equation close to hardware
and allows to solve it with cheap and simple hardware components.\end{abstract}


\global\long\def\atan{\operatorname{atan}}

\global\long\def\atanh{\operatorname{atanh}}

\global\long\def\asin{\operatorname{asin}}

\global\long\def\asinh{\operatorname{asinh}}

\global\long\def\acos{\operatorname{acos}}

\global\long\def\acosh{\operatorname{acosh}}

\global\long\def\floor{\operatorname{floor}}

\global\long\def\Kc{K_{\mathrm{c}}}

\global\long\def\Kh{K_{\mathrm{h}}}

%\date{Received / Accepted}

%\maketitle

\section{Introduction}

Kepler\textquoteright s equation (KE) is fundamental in many fields
of astrophysics. It relates mean anomaly $M$ and eccentric anomaly
$E$ via the equation
\begin{align}
E-e\sin E & =M(E)\label{eq:KE}
\end{align}
where $M(t)=\tau\frac{t}{P}$ with time $t$ and orbital period $P$.

In practice we often need to solve the inverse the problem $E(M)$.
For instance, in orbit fitting, observing times $t$ are given and
then the location or velocity of an object must be predicted, which
then requires to compute $E$.

Many methods have been proposed to solve KE, such as Newton iterations,
Halley's method, table interpolation, or inverse series \citep{Colwell1993sket.book.....C}.
In \citet[{}][hereafter {\ZeXVIII}]{Zechmeister2018A&A...619A.128Z},
we proposed to use a CORDIC-like algorithm. CORDIC (Coordinate Rotation
Digital Computer) was invented by \citet{Volder1959cordic} and can
compute many elementary functions (e.g. cosine, sine, multiplication)\footnote{We provide an online demo at \url{https://raw.githack.com/mzechmeister/ke/master/cordic/js/cordic.html}.}
and needs only additions and bitshifts. We will briefly review the
CORDIC concept in \autoref{sec:CORDIC-classic}.

In {\ZeXVIII}, the rotation directions are set accurately. However,
already this step requires a multiplication in each iteration. To
overcome this shortcoming, we study here the idea to simply ignore
the scale change in the direction decision, still hoping for a correct
convergence (\autoref{sec:CORDIC-KEdbl}). We will demonstrate that
this approach is indeed purposeful given some appropriate adjustments.
Finally, we discuss an implementation (\autoref{sec:Implementation})
and evaluate the performance of the algorithm (\autoref{sec:Accuracy}).

\section{CORDIC algorithm for elementary functions}
\label{sec:CORDIC-classic}

A complex number can be expressed in Cartesian coordinates 
\begin{align}
z & = x+iy\label{eq:z_cartesian}
\end{align}
as well as in polar coordinates 
\begin{align}
z & = r \exp i\varphi\label{eq:z_polar}
\end{align}
where $x=r\cos\varphi$ and $y=r\sin\varphi$.

When we represent this number $z$ by a sequence of rotations with
angles $\theta_{n}$, these are simple additions in the exponent in
the polar representation or complex multiplications in the Cartesian
representation
\begin{align}
re^{i\sum\theta_{n}}~=~z~ & =~r\prod_{n}(\cos\theta_{n}+i\sin\theta_{n}).\label{eq:z_composite}
\end{align}
Additions are very easy to perform for computers, while multiplications
are usually more expensive, in particular with simple hardware as
years ago. Therefore, \citet{Volder1959cordic} sought to simplify
the product in \autoref{eq:z_composite}. He factored the term $\cos\theta_{n}$
\begin{align}
z & =r\prod_{n}\cos\theta_{n}\prod_{n}(1+i\tan\theta_{n})\label{eq:z_prod_factored}
\end{align}
and allowed only angles of the form
\begin{equation}
\theta_{n}=\sigma_{n}\alpha_{n}
\end{equation}
with 
\begin{align}
\tan\alpha_{n} & =\frac{1}{2^{n-1}} & \text{and} &  & \sigma_{n}\in\{-1,1\}\label{eq:tanan}
\end{align}
for $n\ge1$. So the first angle is $\alpha_{1}=45^{\circ}$ and the
next rotation angles $\alpha_{n}$ are almost halved in each iteration
and the rotation can be clock- or counter-clockwise (positive or negative).
Now, \autoref{eq:z_prod_factored} can be written as

\begin{align}
z & =rK_{N}\prod_{n}\left(1+i\frac{\sigma_{n}}{2^{n}}\right)\label{eq:z_prod_simplified}
\end{align}
where the term
\begin{equation}
K_{N} = \prod_{n}\cos\alpha_{n}=\prod_{n}\frac{1}{\sqrt{1+\tan^{2}\alpha_{n}}}
     = \prod_n \frac{1}{\sqrt{1+4^{-n}}}\label{eq:KN}
\end{equation}
is called scale correction. The factor $K_{N}$ can be pre-computed,
because the $\cos$-function is symmetric and therefore independent
of $\sigma_{n}$ and the absolute values of the rotation angles $|\theta_{n}|=\alpha_{n}$
are pre-defined ($\Kc\equiv K_{\infty}\approx0.607\thinspace253$)\footnote{\label{fn:Pochhammer}The $q$-Pochhammer symbol is defined as $(a;q)_{n}=\prod_{k=0}^{n}(1-aq^{-k})$.
Thus the product series $\Kc^{2}=\prod_{k=0}^{^{\infty}}(1+4^{-k})$
is the special case $(-1;\frac{1}{4})_{\infty}\approx2.71182$. Likewise,
in hyperbolic mode there occurs $\prod_{k=1}^{^{\infty}}(1-4^{-k})=(\frac{1}{4};\frac{1}{4})_{\infty}\approx0.68854$,
which is also a special case of the Euler product.}.

Due to the angle choice in \autoref{eq:tanan}, the remaining product
term can be computed efficiently. This is easier to explain when explicitly
writing an adjacent rotation for the real and imaginary part of $z$
as
\begin{align}
x_{n+1} & =x_{n}-m\frac{\sigma_{n+1}y_{n}}{2^{k_{n+1}}}\label{eq:x_rot}\\
y_{n+1} & =y_{n}+\phantom{m}\frac{\sigma_{n+1}x_{n}}{2^{k_{n+1}}},\label{eq:y_rot}
\end{align}
where the coordinate parameter $m$ is $1$ for the circular case
($-1$ for the hyperbolic and 0 for the linear case).

The multiplication with $\sigma_{n}$ is just a negation in case $\sigma_{n}=-1$.
The multiplication by an integer power of two ($2^{-n}$) is also
very easy for a computer. It is a simple bit shift in binary system;
very similar in a decimal system a division by ten is just a left
shift of the decimal point. Therefore, all multiplications are eliminated.
Only the one multiplication $rK_{N}$ in \autoref{eq:z_prod_simplified}
remains; and in case $r=1$ even this multiplication can be saved
\citep{Walther:1971:UAE:1478786.1478840}; the start vector is initialised
with $(x_{0},y_{0})=(K_{N},0)$.

With these basic equations, the CORDIC algorithm can compute the $\sin$e
and cosine function. Given an input angle $\varphi$, we can approach
it in each iteration with the condition
\begin{equation}
\sigma_{n+1}=\begin{cases}
+1 & \sum\theta_{n}<\varphi\\
-1 & \text{else.}
\end{cases}\label{eq:sigma_n}
\end{equation}

The Cartesian representation is propagated simultaneously with the
same rotation directions $\sigma_{n}$ via Eqn.~\ref{eq:x_rot} and
\ref{eq:y_rot}. So when $\sum\theta_{n}\rightarrow\varphi$, then
$x_{n}\rightarrow\cos\varphi$ and $y_{n}\rightarrow\sin\varphi$.
\Autoref{fig:cordic_modes} illustrates this process.

The convergence range can be derived when performing only positive
rotations resulting in $\sum_{n=1}^{N}|\theta_{n}|=\sum_{n}\atan2^{-(n-1)}\rightarrow1.7433=99.88^{\circ}$
for $N\rightarrow\infty$. An initial rotation with $90^{\circ}$,
which needs no scale correction, can extend the range to $189.88^{\circ}$.

It is also possible to calculate $\atan(x,y)$. So given $x$ and
$y$, the angle $\varphi$ of this vector is wanted. In this mode,
called vectoring, the component $y_{n}$ is driven towards zero.

\citet{Walther:1971:UAE:1478786.1478840} generalised the CORDIC algorithm
with a linear and hyperbolic mode allowing to compute multiplication,
division and the functions $\exp$, $\ln$, $\atanh$, and square
root. \Autoref{tab:CORDIC-modes} gives an overview of the different
modes and \autoref{tab:CORDIC-function} lists the required input
and the corresponding output to obtain various elementary functions.
This diversity demonstrates the capability of this simple algorithm.

However, it must be noted that the hyperbolic mode needs specific
iterations to be done twice for\footnote{The sequence is related to \url{http://oeis.org/A003462}.}
\begin{equation}
k_{n}\ \in\ 4,13,40,121,...,K,3K+1\ =\ \frac{3^{m+2}-1}{2}.\label{eq:kn_repeated}
\end{equation}
This compensates the accumulating problem that subsequent rotation
angles are a little smaller than half, $2\alpha_{n+1}<\alpha_{n}$
(while the circular modes has here some redundancy $2\alpha_{n+1}>\alpha_{n}$,
and the linear mode is exact $2\alpha_{n+1}=\alpha_{n}$). With this
sequence the convergence is overall. From now on, we use the variable
name $n$ for iteration number and $k_{n}$ for the shift sequence.

\begin{figure}
\includegraphics[width=1\linewidth]{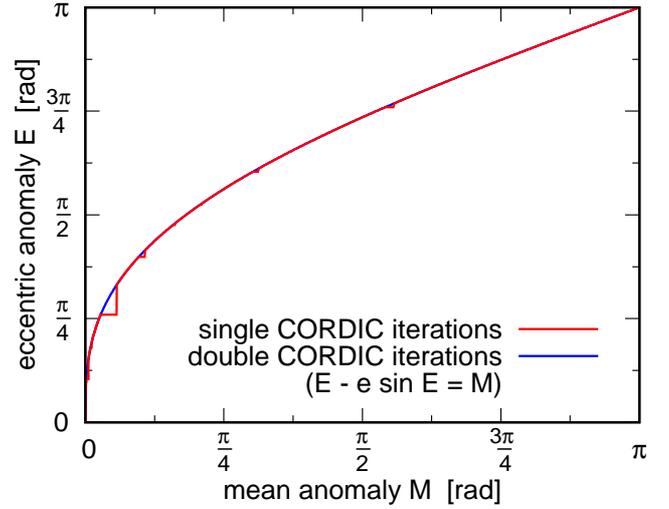}

\caption{\label{fig:ke_single_iter}CORDIC on KE (for $e=1$) with single iterations
(red, shift sequence $k_{n}=0,1,2,3,4,...$) and with double iterations
(blue, $k_{n}=1,1,2,2,3,3,...$).}
\end{figure}

\section{CORDIC double iterations for Kepler's equation}
\label{sec:CORDIC-KEdbl}

To apply CORDIC to Kepler's equation, we proposed in {\ZeXVIII} the
modified condition for the rotation direction\footnote{For reasons of uniformity with code implementation, we list the positive
case first compared to {\ZeXVIII}.}
\begin{equation}
\sigma_{n+1}=\begin{cases}
+1 & E_{n}-e\sin E_{n}<M\\
-1 & \mathrm{else.}
\end{cases}\label{eq:sgn-Z18}
\end{equation}
For readability and with respect to \autoref{eq:KE}, we renamed $\sum\theta_{n}$
with $E_{n}$ and $\varphi$ with $M$ compared to \autoref{sec:CORDIC-classic}.

The decision in \autoref{eq:sgn-Z18} is exact within the working
precision. However, the term $\sin E_{n}$ is not accessible in true
CORDIC, because of the scale change. And a simultaneous compensation
would require a multiplication in every iteration. When the start
vector $E_{0}=0$ is pre-scaled with $K_{N}$, i.e. 
\begin{align}
x_{0} & =K_{N}e\cos E_{0}=K_{N}e\label{eq:x0}\\
y_{0} & =K_{N}e\sin E_{0}=0,\label{eq:y0}
\end{align}
then the term $y_{n}$ converges towards $e\sin E_{n}$ for $n\rightarrow N$.
But at iteration $n$, the relation is 
\begin{align*}
y_{n} & =\frac{K_{n}}{K_{N}}e\sin E_{n}.
\end{align*}
Therefore, $y_{n}$ and $e\sin E_{n}$ deviate by the factor $\frac{K_{n}}{K_{N}}=\prod_{n}^{N}(1+4^{-k_{n}})^{-1/2}$
(in double precision it is negligible for $k_{n}\ge27$).

In this work, we simply propose
\begin{equation}
\sigma_{n+1}=\begin{cases}
+1 & E_{n}-y_{n}<M\\
-1 & \mathrm{else.}
\end{cases}\label{eq:sgn}
\end{equation}
This ignores totally the changing scale. Still, we might hope for
a convergence. \Autoref{fig:ke_single_iter} shows what happens for
the extreme case of $e=1$. Many regions seem to converge, but obviously
others do not converge. The approximation leads sometimes to rotations
into wrong directions and the subsequent rotations seem not to overcome
this.

There are other functions that can have similar issues, for instance
the arcsine \citep{Muller2006}. \citet{Baykov1972} solved the problem
with double iterations, meaning each iteration is executed twice.
In \autoref{subsec:arcsine}, we discuss the arcsine function.

We continue investigating the approach of \citet{Baykov1972}, because
it does not require any modifications of the CORDIC algorithm besides
the sequence for $k_{n}$. We also remind that the hyperbolic mode
needs specific iterations to be repeated, too. So double iteration
is an established workaround. Indeed it turns out, that the double
iterations also work for Kepler's equation (\autoref{fig:ke_single_iter}).

\begin{figure}
\includegraphics[width=1\linewidth]{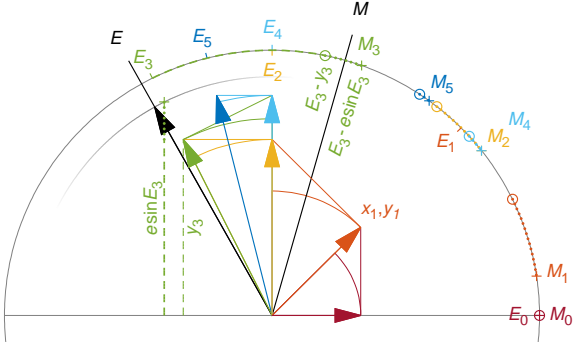}

\caption{\label{fig:ke_vector}Evolution of the vector $x_{n},y_{n}$ (colour-coded
arrows), when dropping the scaling term $\cos\alpha_{n}$ and using
double rotation. The example is for $e=0.9$ and $M=2.08-0.9\sin2.08\approx1.294$
(thus $E=2.08$). With appropriate prescaling by $K_{N}$, the vector
approaches length $e$ (grey arc) after $n=N$ (black arrow) and converges
towards the $E$. The discrepancy between exact intermediate mean
anomaly $M_{n}=E_{n}-e\sin E_{n}$ (crosses) and approximated mean
anomaly $E_{n}-y_{n}$ (open circles) is colour-coded with dotted
arcs on the grey unit circle. Iteration $n=1$ and $n=2$ (or three
and four) have the same angle $\alpha_{n}$.}
\end{figure}
We can explain the success as follows. As already mentioned, $y_{n}$
is a good approximation of $e\sin E_{n}$. A rotation into a wrong
direction can occur, when the intermediate angle is already close
to the target value (see $n=3$ in \autoref{fig:ke_vector}). Then
the true $M_{n}=E_{n}-e\sin E_{n}$ and the approximation $E_{n}-y_{n}$
may lay on different sides with respect to the target $M$ (see \autoref{fig:ke_vector},
a positive rotation $\sigma_{4}=+1$ would be needed according to
$M_{3}$, but $E_{3}-y_{3}$ suggests $\text{\ensuremath{\sigma_{4}=-1}}$).
A wrong rotation moves away by at least $\alpha_{n}$ and needs to
be compensated by the subsequent rotations. In case of single rotations,
all subsequent rotations $\sum_{n+1}\alpha_{n}\approx2\alpha_{n+1}$
can recover $\alpha_{n}$. But the small redundancy in single rotations
(in $m=1$) is generally insufficient to compensate yet the initiating
departure. Double rotations, however, introduce redundancy, which
is the key to overcome the convergence problem, but also the price
to be paid for the approximation.

\section{Hyperbolic mode}
\label{sec:CORDIC-KEh}

Analogous to {\ZeXVIII}, we study whether our double iteration algorithm
is also applicable to the hyperbolic Kepler's equation (HKE)

\begin{equation}
M=e\sinh H-H,\label{eq:KEh}
\end{equation}
where $e\ge1$. Replacing the trigonometric terms by the hyperbolic
analogues, Eqn.~\ref{eq:tanan} and \ref{eq:sgn} become
\begin{align}
\tanh\alpha_{n} & =\frac{1}{2^{k_{n}}}\label{eq:alphah_n}\\
\sigma_{n+1} & =\begin{cases}
+1 & y_{n}-H_{n}<M\\
-1 & \text{else}.
\end{cases}
\end{align}
With $m=-1$ in \autoref{eq:x_rot}, hyperbolic rotations are performed.
The hyperbolic iterations return $x_{N}=e\cosh H_{N}$ and $y_{N}=e\sinh H_{N}$.
The double iterations cover a range of $|H|<2.111$ (in {\ZeXVIII}:
$|H|<4\ln2=2.772$). The scale correction is $K_{\mathrm{h,dbl}}\approx K_{\mathrm{h},\infty}^{2}(1-4^{-4})\approx1.452\,35$.
As suggested in {\ZeXVIII}, large mean anomalies can be handled with
appropriate start values
\begin{align}
H_{0} & =k_{0}\ln2\label{eq:H0}\\
x_{0} & =Ke\cosh H_{0}=\frac{eK}{2}[\exp H_{0}+\exp(-H_{0})]\\
 & =eK\thinspace[2^{k_{0}-1}+2^{-k_{0}-1}]\\
y_{0} & =Ke\sinh H_{0}=\frac{eK}{2}[\exp H_{0}-\exp(-H_{0})]\\
 & =eK\thinspace[2^{k_{0}-1}-2^{-k_{0}-1}].
\end{align}
where the integer $k_{0}$ is taken from
\begin{equation}
k_{0}=\sign M\cdot\max\left[0,\floor\left(1+\log_{2}\left|\frac{M}{e}\right|\right)\right].\label{eq:m}
\end{equation}
We remark that, the start triple requires only additions and bitshifts
and the one multiplication in $eK$. For $k_{0}=0$, the start triple
is similar to the elliptic case ($H_{0}=0$, $x_{0}=1$, and $y_{0}=0$).
For $k_{0}\ne0$, the triple yields a range extension. (A range reduction
as in the circular case, where $E_{0}$ and $y_{0}$ becomes zero,
is not possible.)

\section{Implementation and variants}
\label{sec:Implementation}

We have indicated in the previous section that our algorithm can solve
Kepler's equation. Here we comment about some details of the implementation.
In particular, we briefly explain the properties of fixed-point and
floating-point representation and the consequences for the algorithm.
The discussion brings us close to the basics of computer architecture.

\begin{table}
\centering

\caption{\label{tab:fixpoint}Examples for bit operations (arithmetic shift,
xor, and) in two's complement with 8 bits.}

\begin{tabular}{@{}cccr@{\,}r@{\,}r@{~=~}r@{}}
\hline 
\hline input & operation & output & \multicolumn{4}{c}{decimal expression}\tabularnewline
\hline 
\texttt{0000\,0001} & \verb_<<_2 & \texttt{0000\,0100} & 1 & $\times$ & $2^{2}$ & 4\tabularnewline
\texttt{1111\,1111} &  & \texttt{1111\,1100} & $-1$ & $\times$ & $2^{2}$ & $-4$\tabularnewline
\texttt{0000\,1011} &  \verb_>>_2 & \texttt{0000\,0010} & 11 & $/$ & 4 & 2\tabularnewline
\texttt{1111\,0101} &  & \texttt{1111\,1101} & $-11$ & $/$ & 4 & $-3$\tabularnewline
\texttt{0111\,1111} & \verb_^_$(-1)$ & \texttt{1000\,0000} & $127$ & $\wedge$ & $(-1)$ & $-128$\tabularnewline
\texttt{0111\,1111} & \verb_^_$0$ & \texttt{0111\,1111} & $127$ & $\wedge$ & 0 & $127$\tabularnewline
\texttt{1111\,0101} & \verb_&_$(-1)$ & \texttt{1111\,0101} & $-11$ & $\&$ & $(-1)$ & $-11$\tabularnewline
\hline 
\end{tabular}
\end{table}

\subsection{\label{subsec:fixpoint}Fix-point implementation}

CORDIC was originally invented for systems with fixed-point numbers.
In fixed-point, a float number is mapped linearly into a chosen integer
range. A 64\,bit system can represent about $2^{64}\approx1.8\cdot10^{19}$
numbers. The location of the virtual ``binary point'' depends on
the convention for the mapping function.

Most systems operate with two's-complement arithmetic. That means,
for signed integers the leading bit is preserved to distinguish positive
and negative numbers. The bit is set for negative numbers. In particular,
$-1$ is represented by setting all bits (\autoref{tab:fixpoint},
second row). There is one more negative number than positive numbers
(e.g. $-128$ vs. $127$ in 8\,bit systems).

A multiplication or a division by a power of two ($2^{n}$) is done
quickly by a bitshift to the left or right, respectively. (Similarly,
in decimal system a division by $10^{n}$ is just a left shift of
the decimal point). This is illustrated in \autoref{tab:fixpoint}.
A right shift of the digits performs a division with a round-down
(floor). Therefore, it should not surprise that, for instance, $-11//4=-3$.

The multiplication with $\pm1$ can be implemented in software with
a conditional addition/subtraction via an \lstinline!if!-statement.
A branchless alternative is implemented in \autoref{lst:KE} and \autoref{lst:HKE}.

In both elliptic and hyperbolic case, we can set the binary point
between the 62th and 61th bit. The most significant bit (number 64th)
is used as sign bit. Then next two bits (63 and 62) can represent
$2^{1}=2$ and $2^{0}=1$. This together with the fractional bits
covers a range of $\pm(4-2^{-61})$ and thus includes the convergences
range of $\pm\frac{\tau}{2}$ and $\pm2.111$, respectively.

\subsection{\label{subsec:ieee}Floating-point implementation}

Nowadays, float numbers on desktop computers are usually represented
in floating-point format as specified in IEEE 754 \citep{IEEE2008}.
Here the first bit is the sign bit, followed by the exponent bits
and finally the mantissa.

One possibility to apply CORDIC is to simply convert the float numbers
to fixed-point representation and to continue with \autoref{subsec:fixpoint}.
This concept is realised in \autoref{lst:KE} and in \autoref{lst:HKE}.

If the algorithm should be still carried out with floating-point format,
one can consider the following circumstances. A multiplication with
$-1$ is just a flip of the sign bit. A division by $2^{k}$ is just
a subtraction of $k$ from the exponent. However, one has to catch
the possibility of an exponent underflow. Also, the addition of two
floating point numbers requires an internal normalisation of the exponent.
Hence addition is not much faster than multiplication and both usually
slower than fix-point addition.

\subsection{\label{subsec:Shift-sequence}Shift and angle sequence}

When we chose as shift sequence $k_{n}=0,0,1,1,2,2,...$ for the circular
mode, then double iterations cover a range of $|E|<\sum_{n=1}^{N}\alpha_{n}\approx199.88^{\circ}$,
that means about twice as wide as in single rotations. The scale factor
becomes $K_{\mathrm{dbl}}\approx K_{\infty}^{2}=0.368\,76$.

Alternatively, one can also start with one scale-free pre-rotation
with $\alpha_{1}=90^{\circ}=\frac{\tau}{4}$ in exchange for two rotations
with $\alpha_{1}=\alpha_{2}=45^{\circ}$ ($k_{1}=k_{2}=0$)
\begin{align}
\sigma_{1} & =\begin{cases}
+1 & E_{0}-e\sin E_{0}<M\\
-1 & \text{else}
\end{cases}\ =\begin{cases}
+1 & E_{0}<M\\
-1 & \text{else}
\end{cases}\\
E_{1} & =E_{0}+\sigma_{1}\frac{\tau}{4}\\
x_{1} & =x_{0}\cos\frac{\tau}{4}-\sigma_{1}y_{0}\sin\frac{\tau}{4}=-\sigma_{1}y_{0}=\,0\\
y_{1} & =y_{0}\cos\frac{\tau}{4}+\sigma_{1}x_{0}\sin\frac{\tau}{4}=\hphantom{-}\sigma_{1}x_{0}=\,\sigma_{1}eK
\end{align}
Combined with the sequence $k_{n}=1,1,2,2,3,3,...$, the convergence
range remains the same and the scale factor becomes $K_{\mathrm{dbl}}\approx(\frac{1}{\cos45^{\circ}}K_{\infty})^{2}=2K_{\infty}^{2}=0.737\,51$.
Obviously, the relative speed profit decreases with number of total
iterations.

\subsection{\label{subsec:accumulation}Accumulation}

\begin{figure}
\begin{centering}
\includegraphics[width=1\linewidth]{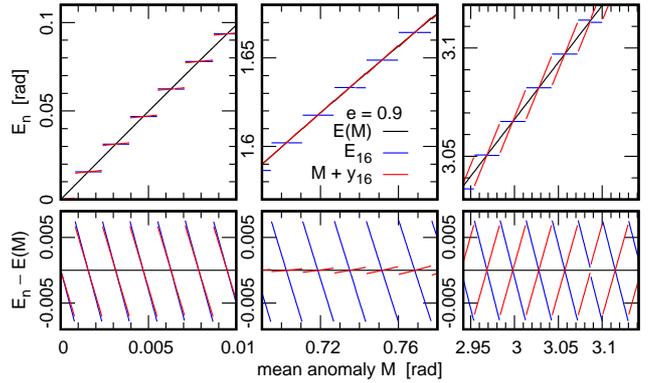}
\par\end{centering}
\caption{\label{fig:bias}{\it Top:} CORDIC output after 16 double iterations ($k_{16}=8$, 29 fractional bits) for eccentricity $e=1$ and three mean anomaly regions. Compared to the exact solution $E$ (black), $E_{16}$ is a step function (blue), while $M+y_{16}$ (red) has slopes. {\it Bottom:} The absolute residuals are smaller than $2^{-7} = 0.0078125$.}
\end{figure}

The condition $E_{n}-y_{n}<M$ in \autoref{eq:sgn} is internally
likely evaluated as $M-E_{n}+y_{n}>0$ and therefore requires two
subtractions (and one comparison). It can be advantageous to reformulate
this as $t_{n}+y_{n}>0$ with $t_{n}=M-E_{n}$ and $t_{0}=M$. This
saves one subtraction (and one variable i.e. memory access) in each
iteration and is possible, because $M$ is a fixed input and $E_{n}$
is needed only in the comparison during the iterations. This accumulation
is a common practice in CORDIC algorithms, where for $e=0$ (so $y_{n}=0$)
the condition $t_{n}>0$ remains. At the end the eccentric anomaly
can be recovered with $E_{N}=M+y_{N}$.

\Autoref{fig:bias} shows that the output $M+y_{N}$ differs a bit from $E_{N}$, but both are within the nominal limits. Positive and negative residuals appear balanced with no visible bias for both cases; a property related to the two-sided design of the proposed CORDIC algorithm (see also Fig. 3 of \ZeXVIII{}). Contrary, one-sided algorithms will be biased. Such a variant was outlined in \ZeXVIII[ (Eq. (13))]{}.

\section{Accuracy and performance study}
\label{sec:Accuracy}

\begin{figure}
\begin{centering}
\includegraphics[width=1\linewidth,trim={0 27 0 0},clip]{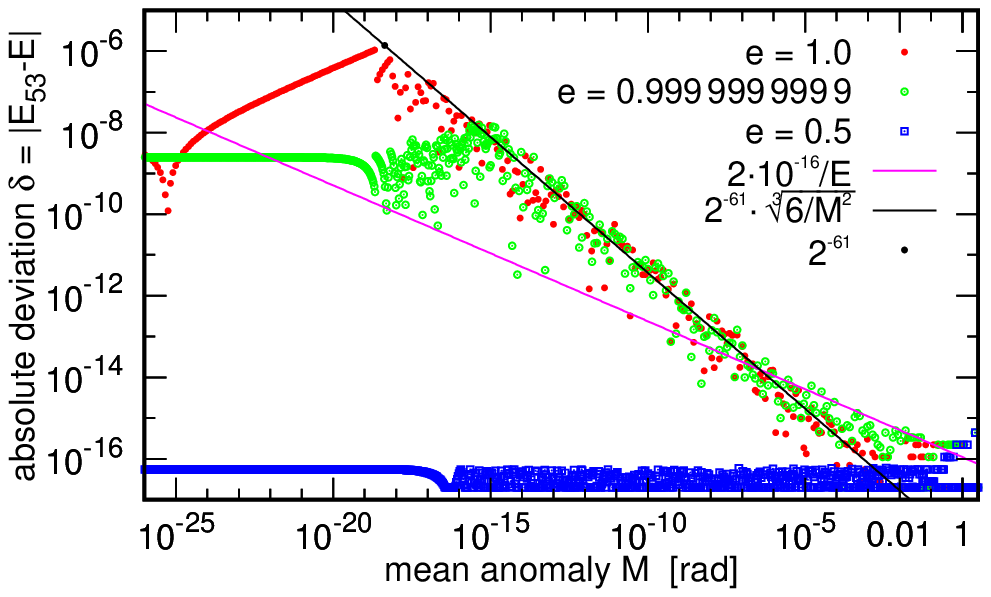}
\par\end{centering}
\begin{centering}
\includegraphics[width=1\linewidth]{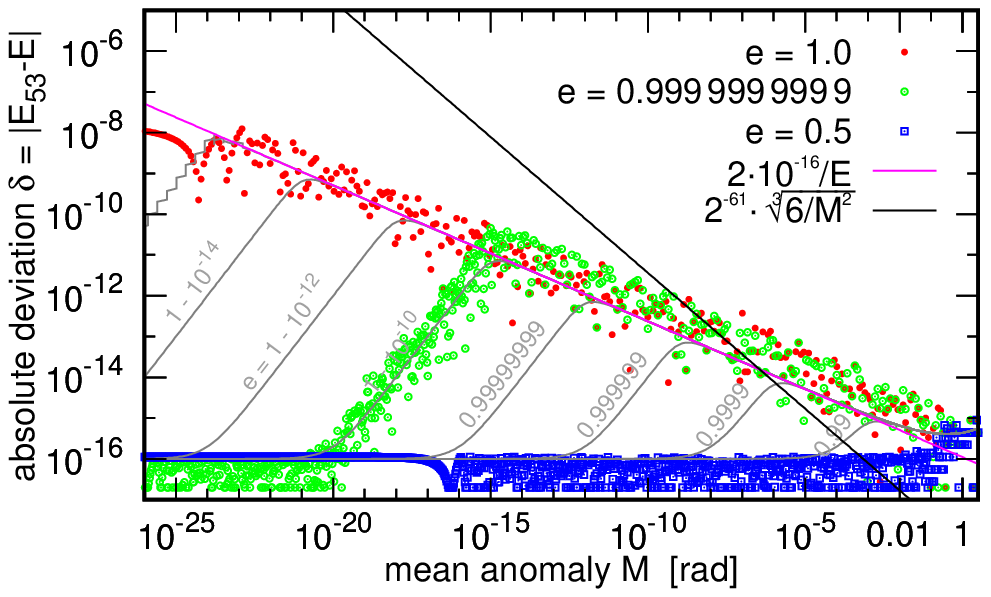}
\par\end{centering}
\caption{\label{fig:accuracy}Accuracy for the fix-point (\emph{top}) and floating-point
algorithm (\emph{bottom}). For visibility, zero deviations ($\delta=0$)
were lifted to $\delta=2\cdot10^{-17}$.}
\end{figure}

\begin{figure}
\begin{centering}
\includegraphics[width=1\linewidth]{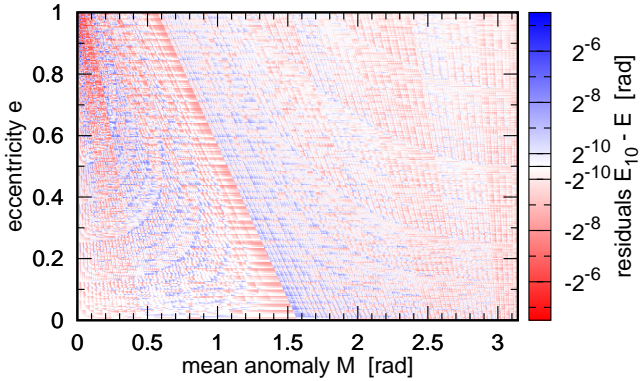}
\par\end{centering}
\caption{\label{fig:map}Residual map for a fix-point algorithm with 10 fractional bits and iterated to the last bit ($k_N=k_{17}=10$).
The colour-coding is log-symmetric.}
\end{figure}

\subsection{\label{subsec:Accuracy-fix}Accuracy of the fix-point algorithm}

We forward calculated with \autoref{eq:KE} 1\,000 $(M(E),E)$ pairs,
with $M$ sampled log-uniformly over $[10^{-26},\frac{\tau}{2}]$.
Here $E$ might be seen as the true value. Then we injected $M$ into
our algorithms to solve the inverse problem $E(M)$. The top panel
of \autoref{fig:accuracy} shows the dependency of the accuracy as
function of $M$ and $e$ for \autoref{lst:KE}. The accuracy becomes
critical in the so called corner of KE at $M=0$ for $e=1$. Here
the function behaves likes a cubic root $E\simeq\sqrt[3]{6M}$ and
the derivative becomes infinite. When using $61$ bits for the binary
fraction (\autoref{subsec:fixpoint}), the step size is $2^{-61}=4.3\cdot10^{-19}$\,rad.
This is the resolution for $M$. The value of the eccentric anomaly
is  $\Delta E=E(M=\Delta M)=\sqrt[3]{6\cdot4.3\cdot10^{-19}}=1.4\cdot10^{-6}$.
This point marks about the largest error and is indicated in the figure.

The general error relation is $\mathrm{d}E=\frac{1}{1-e\cos E}\mathrm{d}M$,
which follows from \autoref{eq:KE}. For $e=1$, it becomes $\mathrm{d}E\simeq\frac{1}{3}\sqrt[3]{\frac{6}{M^{2}}}\mathrm{d}M$
and since $\Delta M$ is constant, the errors declines as $\propto M^{-2/3}$.
The residuals matches those theoretical limits and thus validates
our implementation.

We remind, that \autoref{fig:accuracy} is an extreme magnification
of the corner. The errors decrease quite quickly with eccentricity.
The increase of error for $M\gtrapprox0.1$ should be related to the
simplified conversion between floating-point and fix-point, which
was used for the preparation of the look-up table and the input data.
The mantissa of 64 bit floating-point numbers holds only 53 bits.

Finally, we remark the need for some spare bits. This is investigated here with a short bit system. We divided the range from 0 to 4\,rad into $2^{12}=4096$ mean anomalies having thus 10 fractional bits ($2^{-10})$. Then we limited our fix-point algorithm also to 10 fractional bits (cf. \autoref{lst:KE-12} in \autoref{lst:KE}) and iterated until the last bit, i.e. $k_N=10$ ($N=17$). Still, the residuals in \autoref{fig:map} are overall limited to $2^{-6}$ (with a slight bias towards more negative deviations) for this short bit algorithm. So sometimes the last four iterations yield no improvement. Therefore, some trailing bits ($\sim \log_2 k_N$) are advisable to buffer accumulating truncation errors.

\subsection{\label{subsec:Accuracy-float}Accuracy of the floating-point algorithm}

Overall, the error behaviour of the floating-point algorithm with
double iteration (\autoref{fig:accuracy}) is very similar to the
CORDIC-like algorithm from {\ZeXVIII[ (Fig.~6)]}, who already explained
the functional forms arising for the different eccentricities. Compared
to the fix-point algorithm, the floating-point algorithms have a better
accuracy in the corner, where they profit from the higher precision
to represent small numbers. On the other hand, the fix-point version
has better accuracy at about $M\gtrsim10^{-5}$.

The algorithm used for \autoref{fig:accuracy} employed the shift
sequence starting with $k_{n}=0,0,1,1,2,2,...$ and without accumulation.
For small mean anomalies, the rotation will go back and forth to zero
in the first iterations, thus keeping properties of the one-sided
algorithms in {\ZeXVIII}. However, the two optimisations suggested
in \autoref{subsec:Shift-sequence} and \autoref{subsec:accumulation}
will worsen the accuracy to a level comparable to the fix-point algorithm,
because for floating-point the order of additions does matter (while
in fix-point format this does not affect the precision). Both cases
lead to the additions of small numbers to big numbers, and thus floating-point
numbers cannot utilise the better representation of small numbers.

\subsection{Performance}
\label{subsec:Performance}

\begin{figure}
\begin{centering}
\includegraphics[width=1\linewidth]{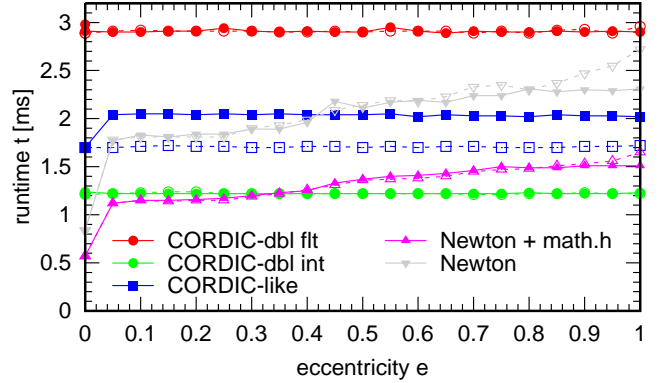}
\par\end{centering}
\caption{\label{fig:runtime}Execution time as function of eccentricity for
various algorithms. For solid curves, $M$ was distributed uniformly,
for dashed curves $E$.}
\end{figure}
We have shown that Kepler's equation can indeed be solved with the
shift-and-add algorithm CORDIC using minor modifications. Thus, the
pressing question is: how fast is the algorithm?

We implemented the algorithms (as well as the comparison solvers)
in plain C and measured with the Python's \texttt{timeit} feature
the execution times to solve \autoref{eq:KE} 10\,000 times and for
21 eccentricities between 0 and 0.999\,999. The maximum shift index
in the CORDIC versions was $k_{N}=28$ corresponding to a precision
of $2^{-28}=3.7\cdot10^{-9}$. The mean anomalies $M$ were distributed
uniformly between $0...180^{\circ}$ (case $\unif(M)$). In a second
test, mean anomalies were generated, whose eccentric anomalies $E$
were distributed uniformly (case $\unif(E)$). This gives some more
weight to small mean anomalies and shall probe whether there is a
dependency on $M$, since some solvers are slower in the corner.

\Autoref{fig:runtime} shows the results. The CORDIC double rotation
algorithms performed $N_{\mathrm{dbl}}=56$ iterations\footnote{For simplicity, we just truncated the shift sequence of the $k_{N}=61$
version, which performs double iterations until $k\text{\ensuremath{\le}}26$,
thus $N_{\mathrm{dbl}}=2\cdot27+(28-26)=56$. But already for $k>14$,
scale corrections are smaller than $4^{-14}=2^{-28}$. Thus $N_{\mathrm{dbl}}=2\cdot15+(28-14)=44$
would be sufficient, too, promising a 21\,\% shorter runtime.}. The runtime is independent of eccentricity and the mean anomaly.
The fix-point version is about 2.2 times faster than the floating-point
version. Thus there is a noticeable benefit. However, it is also ``only''
a factor of two, since the floating-point code needs two multiplications
in each iterations. This shows that multiplications have been optimised
over the last years in current computer processing unit (CPU), crowded
out CORDIC algorithms from those architectures.

For comparison, we show the CORDIC-like version from {\ZeXVIII},
which is a floating-point algorithm. We used a one-sided variant,
which is faster for small $E$. This may partly explain the performance
increase for the case $\unif(E)$. Moreover, this version used an
\lstinline!if!-branch, thus branch prediction may affect the results,
too. In any case, since the CORDIC-like version needs only $N=29$
iterations, it is faster than its double iteration companion. Yet,
the new fix-point version is even 1.5 times faster.

Finally, we see that the speed of the fix-point version compares well
with Newton's method, which used the start guess $E_{0}=M+0.85e$.
The cosine and sine functions were called from the standard \texttt{math.h}
library. Since the source of their implementation can be hard to track
down, a self-programmed sine and cosine functions were tested as well.
The runtime increases with eccentricity. At $e=1$, the $\unif(E)$
case take a bit longer than the $\unif(M)$ case, which indicates
slower convergence in the corner (at $M\approx0$).

The performed tests can only serve as an orientation. There are many
aspects, which can alter the outcome. Furthermore, the algorithm in
{\ZeXVIII} yields as output also the terms $\cos E$ and $\sin E$,
which will be needed for subsequent computation. Our CORDIC algorithm
delivers $e\cos E$ and $e\sin E$. In case division by $e$ appears
disadvantageous in particular for small eccentricities, one can recompute
the terms from $E$ or alternatively propagate them in parallel with
the algorithm.

\section{Further discussion}

\subsection{Unifying the Keplerian CORDIC modes}

CORDIC has three coordinate systems: linear, circular, and hyperbolic.
We have extended here the circular and hyperbolic mode to the elliptic
and hyperbolic case of Kepler's equations. Thus, the question is nearby,
whether there is an analog extension for the linear mode. To address
this, we summarise the main difference in the input and the direction
decision of the various modes
\begin{align}
 & \mathbb{\text{rotating}} & t_{n}\phantom{{+}y_{n}} & >0 & t_{0} & =\varphi\label{eq:cond_rot}\\
 & \mathbb{\text{vectoring}} & -y_{n} & >0 & t_{0} & =0\label{eq:cond_vec}\\
 & \text{arcsine} & -\varphi+y_{n} & >0 & t_{0} & =0\label{eq:cond_asin}\\
 & \text{KE} & t_{n}+y_{n} & >0 & t_{0} & =M\label{eq:cond_KE}\\
 & \text{HKE} & -t_{n}-y_{n} & >0 & t_{0} & =-M,\label{eq:cond_HKE}
\end{align}
which includes the arcsine mode for completeness. The Keplerian modes
appear as mixture of rotating and vectoring, and we call it ``keplering''.

Rotation and vectoring handle the different coordinate systems via
the parameter $m$, which appears in \autoref{eq:x_rot}, but not
in Eqn.~\ref{eq:cond_rot}--\ref{eq:cond_asin}. Now, to unify the
Keplerian modes, we can suggest to introduce $m$ in Eqn.~\ref{eq:cond_KE}
and \ref{eq:cond_HKE} as
\begin{align}
 & \text{GKE} & m(t_{n}+y_{n}) & >0 & t_{0} & =mM.
\end{align}

For the linear mode with $m=0$, this unification appears pointless,
as the outcome will not depend on $M$ at all. Well, if we associate
the linear Keplerian mode with the case of \emph{radial trajectories},
it would indeed complete the picture. KE and HKE solve for a time
dependent auxiliary angle, but in radial cases the angle is fixed
and time independent.

Another possibility for unification are base angles~$m\alpha_{n}$
(thus negative for HKE) along with the condition $t_{n}+my_{n}>0$
and $t_{0}=M$. Then for $m=0$, all base angle would be zero.

\subsection{Barker's equation}

We also considered that a linear mode extension might be associated
with parabolic orbits. This special case is handled with Barker's
equation \citep{Colwell1993sket.book.....C}, which is given by
\begin{equation}
M=D+\frac{1}{3}D^{3},\label{eq:Barker}
\end{equation}
where $D$ is an auxiliary variable, similar to the eccentric anomaly
$E$ and $H$, and related to true the anomaly by 
\begin{equation}
\tan\frac{\nu}{2}=D.
\end{equation}

\Autoref{eq:Barker} is a cubic equation, whose explicit solution
is often given as
\begin{equation}
D=B-\frac{1}{B}\label{eq:D_Meire}
\end{equation}
with $B=\sqrt[3]{W+\sqrt{W^{2}+1}}$ \citep{Meire1985JBAA...95..113M}
and $W=\frac{3}{2}M$. Computing this in a CORDIC framework requires
five operations: division, hypotenuse, as well as exp, div, and ln
for the cubic root ($3M$ are simply three additions).

The equivalent solution with hyperbolicus functions
\begin{equation}
D=2\sinh\frac{\asinh\frac{3M}{2}}{3}\label{eq:D_new}
\end{equation}
seems to be less known (\autoref{sec:D_proof}). It requires at most
five CORDIC operations: sinh, div, and, for the arcsine hyperbolicus
(\autoref{eq:asin}), mul, cathetus, and $\atanh$. When using the
double iteration variant for the arcsine hyperbolicus, then the total
costs are about four CORDIC cycles. A further reduction could be done
by optimising the division by three (\autoref{subsec:mul_const}).

Therefore, Barker's equation can be solved with a few nested CORDIC
operations, but we don't see a possibility to tackle it more directly
with CORDIC.

\section{Summary}

In this work, we presented to our knowledge for the first time a shift-and-add
algorithm to solve KE. The features of the algorithm are
\begin{itemize}
\item usage of most basic computer operations (addition, bitshift, xor)
\item small code size and short lookup table
\item independent of math libraries
\item adjustable precision
\item runtime independent of mean anomaly $M$ and $e$.
\end{itemize}
We require only two minor modifications to the normal CORDIC algorithm,
which are the modified direction decisions (\autoref{eq:sgn}) and
repeating each iteration once (double iterations for $k\le\frac{1}{2}k_{N}$).
The modifications constitute a new CORDIC mode, which we call Keplerian
mode or keplering. It is mixture of rotating and vectoring and handles
the eccentric and hyperbolic case of Kepler's equation.

From the perspective of CORDIC, solving Kepler's equation appears
twice as expensive as the computation of a sine function or about
as expensive as the arcsine function, which both are parameter less,
while KE has the parameter $e$.

Albeit we could eliminate all multiplications from the iteration loop,
this is hardly honoured by current desktop computers, which nowadays
have sophisticated multiplier units. Thus CORDIC algorithms have been
displaced from computer processor. While our CORDIC KE solver is well
competitive with Newton's method, its full potential would become
available on architectures, which favour CORDIC methods, such as old
or cheap devices. Also, a wider revival of CORDIC might be possible
in the future.

\section*{Acknowledgements}
I thank Hanno Rein for refereeing the paper, Trifon Trifonov for manuscript reading, and Albert Chan for
helpful discussion about Barker's equation (\autoref{eq:D_new}, \autoref{sec:D_proof}).
This work is supported by the Deutsche Forschungsgemeinschaft under
DFG RE 1664/12-1 and Research Unit FOR2544 \textquotedblleft Blue
Planets around Red Stars\textquotedblright , project no. RE 1664/14-1.

\section*{Data Availability}

No new data were generated or analysed in support of this research.

\bibliographystyle{mnras}
\bibliography{ke_cordic_dbl}

\appendix

\section{Notes on the CORDIC method}
\label{sec:CORDIC_notes}

\Autoref{tab:CORDIC-modes} summarises the general output of CORDIC
for the three coordinate systems $m$ and the two operation modes.
From this one can derive elementary functions using specific inputs
as given in \autoref{tab:CORDIC-function}. For instance, the $\exp$-function
is obtained with $x_{0}=\Kh$ and $y_{0}=\Kh$ in hyperbolic vectoring
(which is even simpler than adding the output of $x_{0}=\Kh$ and
$y_{0}=0$: $\exp x=\cosh x+\sinh x$).

The logarithm is obtained from the atanh-function via the identity\footnote{$\tanh\ln\sqrt{\frac{a}{b}}=\frac{\exp\ln\sqrt{a/b}\ -\ \exp\ln\sqrt{a/b}}{\exp\ln\sqrt{a/b}\ +\ \exp\ln\sqrt{a/b}}=\frac{\sqrt{a/b}-\sqrt{b/a}}{\sqrt{a/b}+\sqrt{b/a}}=\frac{a-b}{a+b}$}
\[
\ln\frac{a}{b}=2\atanh\frac{a-b}{a+b}.
\]
So to compute $\ln a$, one has to set $b=1$, thus $x_{0}=a+1$ and
$y_{0}=a-1$. The final multiplication with two is again a bitshift.
The same mode provides simultaneously the square root~$\sqrt{a}$.
With $x_{0}=a+\frac{\Kh^{2}}{4}$ and $y_{0}=a-\frac{\Kh^{2}}{4},$
the output is directly available in $x_{n}$.\footnote{$\frac{K_{h}^{2}}{4}=0.364512$. Usually, $x_{0}=a+\frac{1}{4}$ and
$y_{0}=a-\frac{1}{4}$ are proposed to get~$\sqrt{a}$. But this
requires a post-multiplication with $\frac{1}{\Kh}$.} The square root is independent of $t$, so this channel can be omitted.

\begin{figure}
\includegraphics[width=1\linewidth]{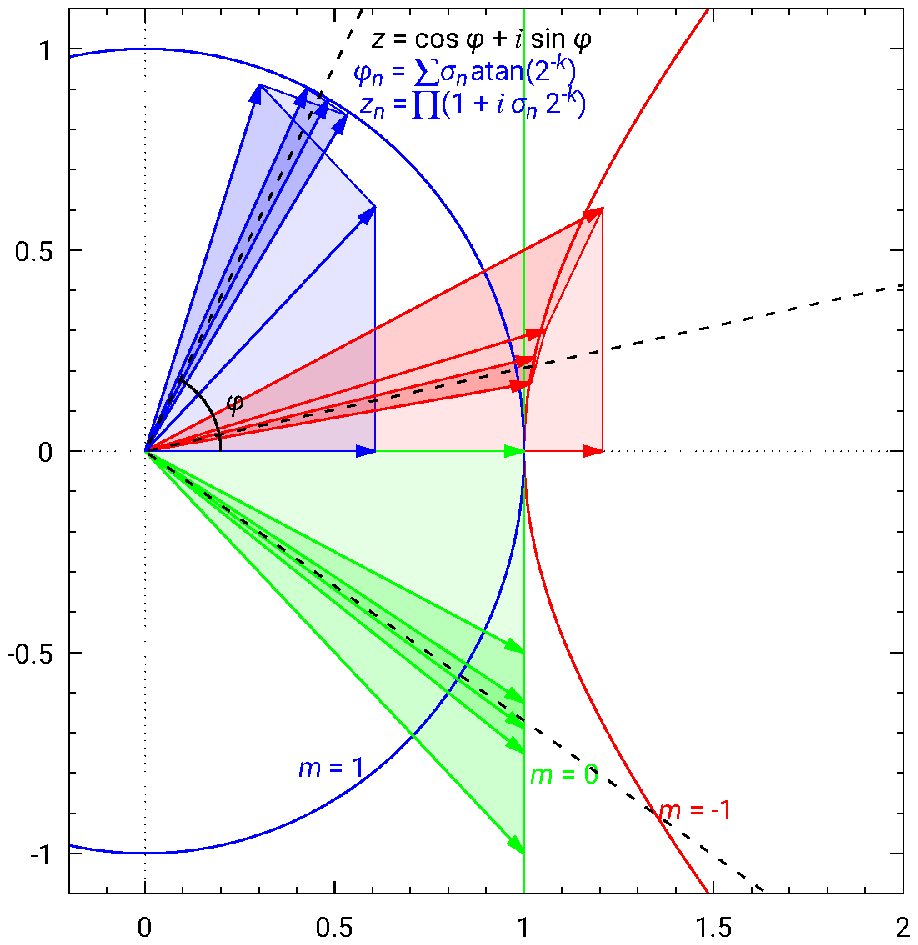}

\caption{\label{fig:cordic_modes}Examples of CORDIC rotations for circular
(blue, $m=1$), linear (green, $m=0$), and hyperbolic mode (red,
$m=-1$). The target angle~$\varphi$ (dashed line) is approached
through rotations with $\alpha_{n}=\protect\atan2^{-k}$. The vectors
change their length with each iteration.}
\end{figure}

\begin{table*}
\caption{\label{tab:CORDIC-modes}General input and output triples for various
CORDIC modes \citep{Walther:1971:UAE:1478786.1478840}.}
\centering%
\newcommand{\lpad}{~~}
\begin{tabular}{@{}llllll@{~~\vline}c@{~~\vline}l@{~\vline}l@{~~\vline}c@{\vline}c@{~~\vline} l@{~\vline}l@{~~\vline}}
\hline 
\hline type & $m$ & $\alpha_{n}$ & $\sum\alpha_{n}$ & $k_{n}$ & \multicolumn{1}{l}{$K_{m,\infty}$} & \multicolumn{3}{c}{rotating ($t\rightarrow0$)} & \multicolumn{1}{c}{} & \multicolumn{3}{c}{vectoring ($y\rightarrow0$)}\tabularnewline
\hline 
\noalign{\vskip-0.5em}
 &  &  &  &  & \multicolumn{1}{l}{} & \multicolumn{7}{c}{}\tabularnewline
\cline{7-7} \cline{9-9} \cline{11-11} \cline{13-13} 
 &  &  &  &  &  & \lpad$x$ &  & \lpad$x$ &  & \lpad$x$ &  & \lpad$x$\vphantom{\LARGE{M}}\tabularnewline
linear & $\hphantom{-}0$ & $2^{-k_{n}}$ & 2 & $0...N$ & 1 & \lpad$y$ & $\ \Rightarrow$ & \lpad$y+xt$ &  & \lpad$y$ & $\ \Rightarrow$ & \lpad0\tabularnewline
 &  &  &  &  &  & \lpad$t$ &  & \lpad0 &  & \lpad$t$ &  & \lpad$t+y/x$\tabularnewline
\cline{7-7} \cline{9-9} \cline{11-11} \cline{13-13} 
\noalign{\vskip-0.5em}
 &  &  &  &  & \multicolumn{1}{l}{} & \multicolumn{1}{c}{} & \multicolumn{1}{l}{} & \multicolumn{1}{l}{} & \multicolumn{1}{c}{} & \multicolumn{1}{c}{} & \multicolumn{1}{l}{} & \multicolumn{1}{l}{}\tabularnewline
\cline{7-7} \cline{9-9} \cline{11-11} \cline{13-13} 
 &  &  &  &  &  & \lpad$\Kc x$ &  & \lpad$x\cos t-y\sin t$ &  & \lpad$\Kc x$ &  & \lpad$\sqrt{x^{2}+y^{2}}$\vphantom{\LARGE{M}}\tabularnewline
circular & $\hphantom{-}1$ & $\atan2^{-k_{n}}$ & 1.743\,286 & $0...N$ & 0.607\,253 & \lpad$\Kc y$ & $\ \Rightarrow$ & \lpad$x\sin t+y\cos t$ &  & \lpad$\Kc y$ & $\ \Rightarrow$ & \lpad0\tabularnewline
 &  &  &  &  &  & \lpad$t$ &  & \lpad0 &  & \lpad$t$ &  & \lpad$t+\atan(y,x)$\tabularnewline
\cline{7-7} \cline{9-9} \cline{11-11} \cline{13-13} 
\noalign{\vskip-0.5em}
 &  &  &  &  & \multicolumn{1}{l}{} & \multicolumn{7}{c}{}\tabularnewline
\cline{7-7} \cline{9-9} \cline{11-11} \cline{13-13} 
 &  &  &  &  &  & \lpad$\Kh x$ &  & \lpad$x\cosh t+y\sinh t$ &  & \lpad$\Kh x$ &  & \lpad$\sqrt{x^{2}-y^{2}}$\vphantom{\LARGE{M}}\tabularnewline
hyperbolic & $-1$ & $\atanh2^{-k_{n}}$ & 1.118\,173 & $1...N$\tablefootmark{*} & 1.207\,497 & \lpad$\Kh y$ & $\ \Rightarrow$ & \lpad$x\sinh t+y\cosh t$ &  & \lpad$\Kh y$ & $\ \Rightarrow$ & \lpad0\tabularnewline
 &  &  &  &  &  & \lpad$t$ &  & \lpad0 &  & \lpad$t$ &  & \lpad$t+\atanh(y,x)$\tabularnewline
\cline{7-7} \cline{9-9} \cline{11-11} \cline{13-13} 
\noalign{\vskip-0.5em}
 &  &  &  &  & \multicolumn{1}{l}{} & \multicolumn{1}{c}{} & \multicolumn{1}{l}{} & \multicolumn{1}{l}{} & \multicolumn{1}{c}{} & \multicolumn{1}{c}{} & \multicolumn{1}{l}{} & \multicolumn{1}{l}{}\tabularnewline
\hline 
\end{tabular}

  \noindent
  \begin{minipage}{\linewidth}
   {\bf Notes.} %\ignorespaces
   \tablefoottext{*}{Specific shift values $k_{n}$ must be repeated, see \autoref{eq:kn_repeated}.}
  \end{minipage}%
\end{table*}

\begin{table}
\caption{\label{tab:CORDIC-function}Input and output triples for some elementary
CORDIC function $m$ (-1 circular, 0 linear, 1 hyperbolic).}

\centering
\renewcommand{\arraystretch}{1.15}

\begin{tabular}{@{}cc@{~}c@{~}ccc@{~}c@{~}c@{}}
\hline 
\hline $m$ & $x_{0}$ & $y_{0}$ & $t_{0}$ & $\Rightarrow$ & $x_{N}$ & $y_{N}$ & $t_{N}$\tabularnewline
\hline 
$\phantom{-}1$ & $\Kc$ & 0 & $\varphi$ & $t\rightarrow0$ & $\cos\varphi$ & $\sin\varphi$ & 0\tabularnewline
$-1$ & $\Kh$ & 0 & $\varphi$ & $t\rightarrow0$ & $\cosh\varphi$ & $\sinh\varphi$ & 0\tabularnewline
$-1$ & $\Kh$ & $\Kh$ & $\varphi$ & $t\rightarrow0$ & $\exp\varphi$ & $\exp\varphi$ & 0\tabularnewline
$\phantom{-}1$ & $1$ & $b$ & 0 & $y\rightarrow0$ & $\frac{1}{\Kh}\sqrt{1+b^{2}}$ & 0 & $\atan b$\tabularnewline
$-1$ & $1$ & $b$ & 0 & $y\rightarrow0$ & $\frac{1}{\Kh}\sqrt{1-b^{2}}$ & 0 & $\atanh b$\tabularnewline
$-1$ & $a+b$ & $a-b$ & 0 & $y\rightarrow0$ & $\frac{2}{\Kh}\sqrt{ab}$ & 0 & $\frac{1}{2}\ln\frac{a}{b}$\tabularnewline
$-1$ & $a+1$ & $a-1$ & 0 & $y\rightarrow0$ & $\frac{2}{\Kh}\sqrt{a}$ & 0 & $\frac{1}{2}\ln a$\tabularnewline
$-1$ & $a+\frac{\Kh^{2}}{4}$ & $a-\frac{\Kh^{2}}{4}$ & 0 & $y\rightarrow0$ & $\sqrt{a}$ & 0 & $\frac{1}{2}\ln\frac{4a}{\Kh^{2}}$\tabularnewline
$\phantom{-}0$ & $a$ & 0 & $\varphi$ & $t\rightarrow0$ & $a$ & $\varphi\cdot a$ & 0\tabularnewline
$\phantom{-}0$ & $a$ & $b$ & 0 & $y\rightarrow0$ & $a$ & 0 & $b/a$\tabularnewline
\hline 
\end{tabular}
\end{table}

For completeness, we list further important functions that can be
derived with additional subsequent CORDIC calls
\begin{align}
\tan x & =\frac{\sin x}{\cos x} & \tanh x & =\frac{\sinh x}{\cosh x}\label{eq:tanx}\\
\cot x & =\frac{\cos x}{\sin x} & \coth x & =\frac{\cosh x}{\sinh x}\label{eq:cotx}\\
\asin x & =\atan(x,\sqrt{1-x^{2}}) & \asinh x & =\atanh(x,\sqrt{x^{2}-1})\label{eq:asin}\\
\acos x & =\atan(\sqrt{1-x^{2}},x) & \acosh x & =\atanh(x,\sqrt{x^{2}+1}).
\end{align}

For instance, $\tan x$ requires one division as a second call, whereas
$\sin x$ and $\cos x$ come simultaneously from circular rotation.
The arcsine is discussed in the next section.

\subsection{The arcsine function}
\label{subsec:arcsine}

\begin{figure}
\begin{centering}
\includegraphics[width=1\linewidth]{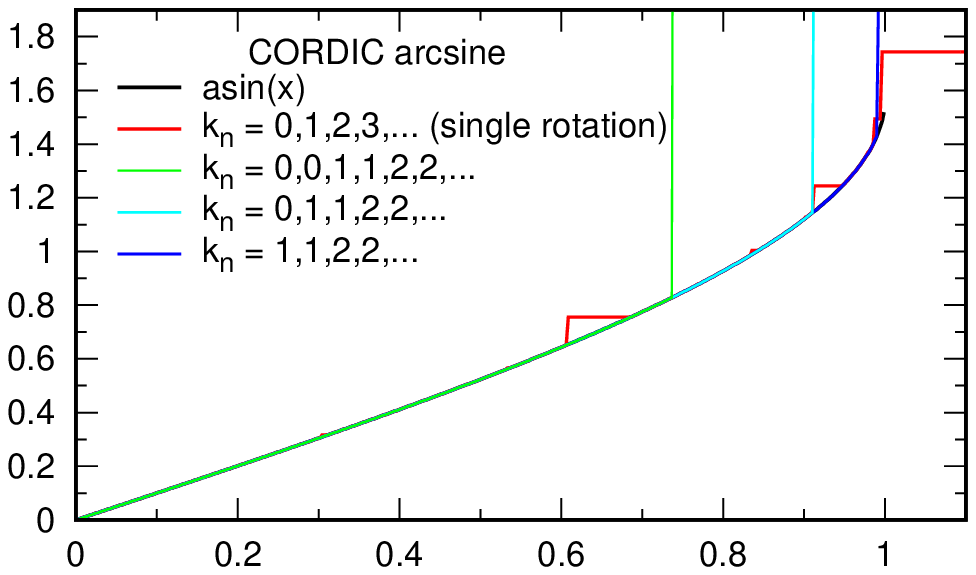}
\par\end{centering}
\caption{\label{fig:arcsine}Output of the arcsine mode. Single rotations (red)
have overall convergence problems. The three double rotation sequences
(green, cyan, blue) converge in a limited range. Full convergence
is achieved by three chained CORDIC operations or double iterations
with quadrant check or scale correction (black, barely visible in
right corner).}
\end{figure}

From \autoref{eq:asin} and \autoref{tab:CORDIC-modes}, it can be
seen that $\asin x$ can be computed with hyperbolic vectoring, which
provides the cathetus $\sqrt{1-x^{2}}$, and circular vectoring, which
executes the $\atan$ function using two arguments and saves the division.
Note, the cathetus needs a multiplicative scale correction (but see
\autoref{subsec:mul_const}), implying three CORDIC calls in total.

A direct way to compute the arcsine is to change the direction decision
as in \autoref{eq:cond_asin} and to drive the $y_{n}$ component
of the vector towards the input argument. However, as explained in
\citet{Muller2006}, the missing scale correction can lead to wrong
decision and erroneous output as illustrated in \autoref{fig:arcsine}.
\citet{Baykov1972} solved the problem with double iterations, meaning
each iterations is executed twice. However, just using the sequence
$k_{n}=0,0,1,1,...$ converges only for $|x|<0.73$, because after
the third rotation the vector cannot recover from an excursion into
an adjacent quadrant. The sequences $k_{n}=0,1,1,2,2,...$ or $k_{n}=1,1,2,2,...$
have a larger convergence range (related to a smaller total scale
corrections). But only an additional quadrant check ($x_{n}>0$) in
each iteration gives full convergence including the arcsine-corner.
We remark that Kepler's equation is bijective and thus does not suffer
from quadrant confusion.

\citet{Takagi1991} developed an alternative method, which is also
called double-CORDIC iteration and employs an auxiliary variable,
which is scale corrected on the fly. \citet{Lang2000} also uses an
on the fly correction, but do not require double rotations.

In summary, there are different ways to compute the arcsine. They
may require modifications and the total costs correspond to about
2--3 normal CORDIC cycles.

\subsection{\label{subsec:mul_const}Multiplication and division with a constant}

The standard CORDIC algorithm requires a scale correction. The scale
correction does not matter in the output $t_{N}$, where the factor
cancels out in the ratio $y/x$ in all modes. But a pre- or post-scaling
is needed, when using the output $x_{N}$ or $y_{N}$ in the modes
$m=1$ or $-1$.

CORDIC provides multiplication and division, and their execution requires
a full CORDIC cycle. However, if the multiplicator is known in advance
and often needed, the efforts can be reduced.

Let's consider first a division by $3$, which occurs in \autoref{eq:D_new}.
The binary representation of $\frac{1}{3}$ is $0.0101010101_{2}...$,
thus its multiplication can be done as $(2^{-2}+2^{-4}+2^{-6}+2^{-8}+...)$.
The corresponding shift-and-add algorithm (essentially a binary multiplier)
can be seen as a CORDIC simplification, which needs only half iterations,
because the iterations with odd shift values can be omitted, since
the linear mode is scale free. Moreover, it does not need direction
decisions and the $t$ channel, because the multiplicator is encoded
in the shift sequence. The bit shifts can be done in parallel.

Multiplications with other important constants, in particular the
scale correction $\Kc\approx0.100\,110\,110\thinspace111\thinspace010_{2}$
and $\Kh\approx1.001\,101\,010\,001\,111_{2}$, could be implemented
in a similar way.

\section{\label{sec:python-code}Code illustration}

In the following, we document a fix-point implementation of the CORDIC
double iteration. The pure Python code shall illustrate the functionality
and low complexity. Further variants (floating-point) and other programming
languages, in particular more performant, low-level C, are maintained
online\ref{fn:thanks}.

\definecolor{blue}{rgb}{0,0,1.0}
\definecolor{deepblue}{rgb}{0,0,0.5}
\definecolor{deepred}{rgb}{0.6,0,0}
\definecolor{deepgreen}{rgb}{0,0.4,0}
\definecolor{coldigit}{RGB}{176,128,0} 
\definecolor{grey}{RGB}{136,135,134} 

\lstset{
language=python,
basicstyle=\ttfamily,
% numbers would be copy pasted
numbers=left,
numberstyle=\tiny,
numbersep=5pt,
 linewidth=\linewidth,
xleftmargin={0.05\linewidth},
%identifierstyle=\color{blue},
commentstyle=\color{grey},
%otherkeywords={self, ln2, cosh},             % Add keywords here
keywordstyle=\bfseries\color{deepgreen},
keywords=[2]{max,range,int}, keywordstyle=[2]\color{deepgreen},
emph={math},             emphstyle=\bfseries\color{blue},
emph=[2]{i64_Ecs,i64_Hcs},       emphstyle=[2]\color{blue},    % Custom highlighting style
frame=tb,
alsoletter={0,1,2,3,4,5,6,7,8,9,.},
keywords=[5]{0,1,2,4,55,60,0.,1.,0.5},
keywordstyle=[5]\color{coldigit},
literate=%
   {,}{,}1,
%   {.}{{{\color{coldigit}.}}}1
%   {9}{{{\color{coldigit}9}}}1,
   float,
   floatplacement=tbp, 
   abovecaptionskip=-5pt
}
\lstdefinestyle{interfaces}{
   float=tp,
   floatplacement=tbp, 
   }
\renewcommand{\lstlistingname}{{\bf Code}} % Listing->Code, doesn't work in preamble

\begin{lstlisting}[float,floatplacement=tbp,linerange={2-2,5-20,38-46},caption={Fix-point algorithm with CORDIC double iteration for Kepler's equation.},label={lst:KE}]
from math import atan, tau

kmax = 53     # largest shift value
R = 1 << 61   # binary point

ak = []       # atan lookup table
K = 1         # scale correction factor
for k in range(kmax+1):
    ak += [(atan(2**-k), k)]
    if 2*k <= kmax:
        K /= 1 + 4**-k
        ak += ak[-1:]

KR = K * R
i64_ak = [(round(a*R), k) for a,k in ak]

def i64_Ecs(M, e):
    t = M - tau*round(M/tau)
    t = round(t * R)
    x, y = round(KR*e), 0
    for a,k in i64_ak:
        s = t+y >> 63
        t -= s ^ s + a
        x, y = x - (s^s+(y>>k)),\
               y + (s^s+(x>>k))
    return M+y/R, x/R, y/R
\end{lstlisting}

\begin{lstlisting}[float,floatplacement=tbp,linerange={3-7,110-122,140-152},caption={Fix-point algorithm with CORDIC double iteration for hyperbolic Kepler's equation.},label={lst:HKE}]
from math import atanh, frexp, log
ln2 = log(2)

kmax = 53     # largest shift value
R = 1 << 61   # binary point

ahk = []      # atanh lookup table
Kh = 1        # scale correction factor
for k in range(1, kmax+1):
    ahk += [(atanh(2**-k), k)]
    if 2*k <= kmax:
        Kh /= 1 - 4**-k
        ahk += ahk[-1:]

KhR = Kh * R
i64_ahk = [(round(a*R), k) for a,k in ahk]

def i64_Hcs(M, e):
    m = max(0, frexp(M/e)[1])
    eK = round(e * KhR) >> 1
    x = (eK<<m) + (eK>>m)
    y = (eK<<m) - (eK>>m)
    if M < 0: m, y = -m, -y
    t = M + m*ln2
    t = -round(t * R)
    for a,k in i64_ahk:
        s = -t-y >> 63
        t -= s ^ s + a
        x, y = x + (s^s+(y>>k)),\
               y + (s^s+(x>>k))
    return y/R-M, x/R, y/R
\end{lstlisting}

\Autoref{lst:KE} implements the algorithm described in \autoref{sec:CORDIC-KEdbl}.
In this snippet, the largest shift is hardcoded in line \autoref{lst:KE-3}
and thus sets the number of iterations.  The precision depends furthermore
on the location of the binary point (\autoref{subsec:fixpoint})\footnote{Because Python integers have arbitrary precision, calculations beyond
64 bits would be possible by increasing this number.} defined in line \autoref{lst:KE-4}.
The lookup
table (\autoref{lst:KE-6}, \autoref{lst:KE-9}) stores the tuples
of the basis angles and the shift sequence ($\alpha_{n}=\atan\frac{1}{2^{k_{n}}}$,
$k_{n}$). If the correction term $4^k$ in \autoref{eq:KN}
is numerical yet larger than $2^{k_N}$ (\autoref{lst:KE-10}), i.e. $k<=k_N/2$, the required
scale correction is accumulated (\autoref{lst:KE-11}) and the last
table entry is repeated (\autoref{lst:KE-12}). Then the table and the scale factor are converted to fix-point (64
bit integers, \autoref{lst:KE-14}--\autoref{lst:KE-15}). The input
of algorithm are floating-point numbers. Thus after a simple range
reduction in \autoref{lst:KE-18} (note that there are more accurate
range reduction algorithms, \citealt{Payne1983}) and pre-scaling
with $K$, the start vector is converted to fix-point (\autoref{lst:KE-19}--\autoref{lst:KE-20}).
The multiplication $e*K$ (\autoref{lst:KE-20}) is the only true
multiplication (but see \autoref{subsec:mul_const}). (The multiplication
with \lstinline!R! and integer conversion can be done with mantissa
extraction and bit shift).

The CORDIC iterations start in \autoref{lst:KE-21}. The comparison
of \autoref{eq:sgn} is done by an addition and a subsequent arithmetic
right bit shift, which extracts the sign bit (\autoref{lst:KE-22}).
The variable \lstinline!s! is either 0 or -1. The bitwise operation
\lstinline!s^(s+x)! (alternatively \lstinline!(x^s)-s!) modifies
the sign of \lstinline!x! in a branchless fashion. In case of \verb_s=0_
($\sigma_{n}=1$), the variable \verb_x_ is not modified. In case
of \verb_s=-1_ ($\sigma_{n}=-1$), the result is an arithmetic negation
(\verb_-x_, see also \autoref{tab:fixpoint}). This can be directly
implemented in hardware (adder-substractor). Line \autoref{lst:KE-23}
accumulates the angles as described in \autoref{subsec:accumulation}.
Finally, the fix-point numbers are converted back to floating-point
(\autoref{lst:KE-26}).

As an example, calling the function with \texttt{i64\_Ecs(2-sin(2),~1)}
should return the triple \texttt{(2.0, -0.41614683654714246, 0.9092974268256817)}.

The code for the hyperbolic Kepler equation (\autoref{sec:CORDIC-KEh},
\autoref{lst:HKE}) is very similar. It requires the hyperbolic arctangent
(\autoref{lst:HKE-10}). The range extension is done in \autoref{lst:HKE-19}--\autoref{lst:HKE-22}.

\section{Barker's equation}
\label{sec:D_proof}

The equivalence of Eqn.~\ref{eq:D_Meire} and \ref{eq:D_new} follows
from
\begin{align}
B-\frac{1}{B} & =e^{\ln B}-e^{-\ln B}=2\sinh\ln B\\
 & =2\sinh\ln\sqrt[3]{W+\sqrt{W^{2}+1}}=2\sinh\frac{\asinh W}{3}.
\end{align}
The last step employed the identity \citep[e.g.][Eq.~(73)]{Fukushima1997CeMDA..68..121F}
\begin{align}
\asinh x=\ln\left(x+\sqrt{x^{2}+1}\right),\label{eq:ident_asinh}
\end{align}
which itself can be verified with the substitution $x=\sinh t$
\begin{align}
t & =\ln\left(\sinh t+\sqrt{\sinh^{2}t+1}\right)=\ln\left(\sinh t+\sqrt{\cosh^{2}t}\right)\\
 & =\ln\left(\sinh t+\cosh t\right)=\ln e^{t}=t.
\end{align}

\end{document}